\documentclass[preprint,superscriptaddress,secnumarabic,amsfonts,amsmath,amssymb]{revtex4}
\usepackage[dvips]{color}
\usepackage{tabularx}
\usepackage{array}
\usepackage{rotating}
\usepackage{epsfig}
\usepackage{amsmath}
\usepackage{graphicx}

\begin{document}

\title{Interacting Holographic dark energy in chameleon tachyon cosmology }

\author{H. Farajollahi}
\email{hosseinf@guilan.ac.ir} \affiliation{Department of Physics,
University of Guilan, Rasht, Iran}
\author{A. Ravanpak}
\email{aravanpak@guilan.ac.ir} \affiliation{Department of Physics,
University of Guilan, Rasht, Iran}
\author{G. Farpour Fadakar}
\email{gfadakar@guilan.ac.ir} \affiliation{Department of Physics,
University of Guilan, Rasht, Iran}

\date{\today}

\begin{abstract}
We propose in this paper an interacting holographic dark energy (IHDE) model in chameleon--tachyon cosmology by interaction between the components
of the dark sectors. In the formalism, the interaction term emerges from the scalar field coupling matter lagrangian in the model rather than being inserted into the formalism as an external source for the interaction. The correspondence between the tachyon field and the holographic dark energy (HDE)
densities allows to reconstruct the tachyon scalar field and its potential in
a flat FRW universe. The model can show the accelerated expansion of the universe and satisfies the observational data.
\end{abstract}


\maketitle


\section{Introduction}

The universe acceleration, shown
by several astronomical observations, indicates
the existence of a mysterious exotic matter called dark energy (DE) \cite{Riess}. In the classical gravity,  whereas the cosmological constant, $\Lambda$, is the most prominent candidate of DE with the equation of state (EoS) parameter equals -1 \cite{Weinberg}, there are strong evidence for a dynamical DE equation of state. However, the problem of DE, its energy density and EoS parameter is still an unsolved problem in classical gravity and may be in the context of quantum gravity we achieve a more inclusive insight to its properties \cite{Cohen}. The HDE model is an attempt to apply the nature of DE within the framework of quantum gravity \cite{Hsu}. The holographic principle states that the number of degrees of freedom
describing the physics inside a volume (including gravity) is bounded by the
area of the boundary which encloses this volume and thus relates to entropy scales with the enclosing
area of the system \cite{Hooft}. Since the entropy scales like the area rather than the volume, the fundamental degrees of freedom describing the system are characterized by an effective quantum field theory in a box of size $L$ with one fewer space dimensions and with planck-scale UV cut-off $\Lambda$ \cite{Setare5}.

As pointed out by the author in \cite{ Myung}, in quantum
field theory, short distance cut-off $\Lambda$ is related to long distance cut-off $L$ due to the limit
set by forming a black hole. In other words, if $\rho_\Lambda$ is the quantum zero-point energy density caused by a UV cut-off $\Lambda$, the total energy of the system with size $L$
should not exceed the mass of the same size black hole, i.e. $L^3\rho_\Lambda\leq LM_p^2$, where $M_p=\frac{1}{\sqrt{8\pi G}}$ denotes Planck mass \cite{Liu}. For the largest $L$, holographic energy density is given by,
\begin{equation}\label{rol}
\rho_\Lambda = 3c^2M_p^2/L^2,
\end{equation}
in which c is a free dimensionless parameter and coefficient three is for convenience \cite{Mohseni}. Assuming $L$ as the universe current size, the Hubble radius $H^{-1}$, then the DE density is close to the observational result, which we identify with the radius of the future event horizon,
\begin{equation}\label{Re}
R_e=a\int_a^\infty\frac{da'}{Ha'^2},
\end{equation}
where $a$ is scale factor of the universe \cite{Wang}. Choosing the future event horizon as the UV cut-off allows us to construct a satisfactory HDE model for the universe accelerating expansion which presents a dynamical
view of the DE and is consistent with observational
data \cite{Zhang}.

The Interacting HDE (IHDE) models have been widely investigated in last few years \cite{Q}. Among all, the stability analysis of the IHDE solutions and their dynamical characteristics \cite{Setare4}, IHDE due to the bulk-brane interaction \cite{Setare2} and IHDE in non-flat cosmological model \cite{Setare1} are of particular interest.

 In particular, relevant works in the holographic tachyon models in both interacting and non interacting cases can be found in \cite{Bagla}\cite{Setare3}\cite{Setare5}. In here, we would like to extend the previous works carried out in the IHDE models, by studying a chameleon-tachyon cosmological model. In our model the scalar field in the formalism plays two roles; as a chameleon field interacts with the matter in the universe and as a tachyon field plays the role of DE. In the previous works in IHDE models, the interacting term between IHDE and dark matter (DM) is not unique and its introduction in the formalism based on phenomenological grounds in studying a suitable coupling between quintessence scalar field and pressureless cold dark matter (CDM) to alleviate the coincidence problem \cite{Wang}. However, in our work the interacting term naturally appears in the model from the interaction between chameleon field and matter field in the universe in the formalism.

\section{The Model}

We consider the chameleon gravity in the presence of CDM with a tachyon potential in the action
given by\cite{Ravanpak}\cite{Ravanpak2},
\begin{eqnarray}\label{action}
S=\int[\frac{M_p^2R}{2}-V(\phi)\sqrt{1-\partial_\mu \phi \partial^\mu \phi}+f(\phi)\mathcal{L}_{m}]\sqrt{-g}d^{4}x,
\end{eqnarray}
where $R$ is Ricci scalar. Unlike the usual Einstein-Hilbert action, the matter
Lagrangian ${\cal L}_{m}$ is modified as $f(\phi){\cal L}_{m}$, where $f(\phi)$ is
an analytic function of $\phi$. This last term in the Lagrangian brings about the nonminimal
interaction between the matter and chameleon field.
The variation of action (\ref{action})  with respect to the metric tensor components in a spatially flat FRW  cosmology yields the following field equations,
\begin{eqnarray}
3H^{2}M_p^2&=&\rho_{m}f+\frac{V(\phi)}{\sqrt{1-\dot{\phi}^{2}}},\label{fried1}\\
M_p^2(2\dot{H}&+&3H^2)=-\gamma\rho_{m}f+V(\phi)\sqrt{1-\dot{\phi}^{2}},\label{fried2}
\end{eqnarray}
where $ H=\frac{\dot{a}}{a}$. In the above, we also assumed a perfect fluid filled the universe with the EoS $p_{m}=\gamma\rho_{m}$. We can rewrite the above equations as,
\begin{eqnarray}
3H^{2}M_p^2&=&\rho_{m}f+\rho_{tac},\label{fried11}\\
M_p^2(2\dot{H}&+&3H^2)=-p_mf-p_{tac}\label{fried22}
\end{eqnarray}
where $\rho_{tac}$ and $p_{tac}$ are respectively the energy density and pressure of the tachyon. The variation of the action (\ref{action}) with respect to the scalar field  $\phi$ provides the wave
equation for the scalar field as
\begin{eqnarray}\label{phiequation}
\ddot{\phi}+(1-\dot{\phi}^{2})(3H\dot{\phi}+\frac{V^{'}}{V})=\frac{\epsilon f^{'}}{V}(1-\dot{\phi}^{2})^{\frac{3}{2}}\rho_{m},
\end{eqnarray}
where prime "  $^\prime$  " indicated differentiation with respect to $\phi$ and $\epsilon=\frac{1-3\gamma}{4}$.
The scalar field as a tachyon field plays the role of IHDE and by coupling with the matter field as a chameleon field plays the role of DM. Then, by their interaction, the energy densities no longer satisfy independent conservation
laws, rather, they obey:
\begin{eqnarray}
\dot{\rho_{ch}}&+&3H\rho_{ch}=Q
\end{eqnarray}
\begin{equation}\label{ct}
\dot{\rho_{tac}} + 3H(1+\omega_{tac})\rho_{tac}=-Q
\end{equation}
where "ch" stands for chameleon, $\rho_{ch}=\rho_m f$ and $Q = \epsilon\rho_{m}\dot{f}$ is the interaction term. In $Q$, $\dot{f}$ gauges the intensity of the coupling between matter and chameleon field. For $\dot{f}=0$, there is no interaction between DM and IHDE. The $Q$  term measures the different evolution of the DM due to its interaction with the IHDE which gives rise to a
different universe expansion. The remarkable point concerning the interaction term is that in comparison to the other IHDE models where the form of the interaction term $Q$ is not unique and built into the formalism from the outset, in our model the interaction term naturally appears in the model directly as a function of the chameleon coupling function $f(\phi)$ and $\rho_m$ and indirectly as a function of Hubble parameter $H$ and $\rho_{tac}$. For the decay of HDE into DM in the interaction we have to take $f_0$ to be positive similar to the case of $b^2>0$ in the interaction term $Q=b^2H\rho_m$ introduced in other works \cite{Edmund}.

In order to impose the holographic nature to the
tachyon, one should identify the energy density of tachyon field with the energy density of the cosmological constant. From the definition of the holographic energy density, equation (\ref{rol}), and that of the
future event horizon (\ref{Re}) we obtain \cite{Z}--\cite{X},
\begin{equation}\label{rho}
\rho'_{tac}\equiv\frac{d\rho_{tac}}{dx}=-6M_p^2H^2\Omega_{tac}(1-\frac{\sqrt\Omega_{tac}}{c}),
\end{equation}
where the derivative in here and from now on is with respect
to $x=ln (a)$, also, $\Omega_{tac}=\rho_{tac}/(3M_p^2H^2)$ and $\rho_\Lambda=\rho_{tac}$. We can rewrite equation (\ref{ct}) as
\begin{equation}\label{ro1}
\rho'_{tac} + 3(1 + \omega_{tac})\rho_{tac} =\frac{Q}{H}.
\end{equation}
 From equations (\ref{rho}) and  (\ref{ro1}) we drive the dynamical EoS parameter as
\begin{equation}\label{omegat}
\omega_{tac}=-\frac{1}{3} - \frac{2\sqrt\Omega_{tac}}{3c}+ \frac{Q}{9H^3M_p^2\Omega_{tac}}.
\end{equation}
We have to note, however, that although the above EoS parameter attributed to the tachyon field, the presence of the $f(\phi)$ in the interaction term is directly related to the chameleon field and DM. So the situation is somewhat ambiguous due to entanglement of the DE and DM. Under these conditions, one can not easily classify the quintessence or phantom unless the interacting term is very weak according
to observations. Using equations (\ref{ro1}) and (\ref{omegat}) and the definition
of $\Omega_{tac}$ we find
\begin{equation}\label{Hz}
\frac{H'}{H}= -\frac{\Omega'_{tac}}{2\Omega_{tac}} + \frac{\sqrt{\Omega_{tac}}}{c} - 1.
\end{equation}
Also, by replacing $\dot H = H'H$ and $p_{tac} = \omega_{tac}\rho_{tac}$
into the acceleration equation (\ref{fried2}), one obtains,
\begin{equation}\label{Hz2}
\frac{H'}{H}= -\frac{3}{2} +\frac{\Omega_{tac}}{2}+\frac{\Omega_{tac}^{3/2}}{c}-\frac{Q}{6H^3M_p^2}.
\end{equation}
From equations (\ref{Hz}) and (\ref{Hz2}), we drive the evolution
equation for $\Omega_{tac}$ in terms of the redshift $z$ as,
\begin{eqnarray}\label{omegaz}
\frac{d\Omega_{tac}}{dz}&=&-(1+z)^{-1}[\Omega_{tac}(1-\Omega_{tac})(1+\frac{2\sqrt{\Omega_{tac}}}{c})\nonumber\\
 &+&\frac{Q\Omega_{tac}}{3H^3M_p^2}],
\end{eqnarray}
which controls the dynamics of the IHDE model. From equation (\ref{Hz2}), the evolution of the Hubble parameter
$H(z)$ with respect to the redshift $z$ can be expressed as
\begin{equation}\label{Hz3}
\frac{dH}{dz}= -(1+z)^{-1}H(-\frac{3}{2}+\frac{\Omega_{tac}}{2}+\frac{\Omega_{tac}^{3/2}}{c}-\frac{Q}{6H^3M_p^2}).
\end{equation}
Using equation (\ref{fried11}), and $\Omega_m=\Omega_{m0}H_0^2H^{-2}(1+z)^3$, we derive the interacting
holographic tachyon potential,
\begin{equation}\label{vphi}
\frac{V(\phi)}{\rho_{cr,0}}= \frac{\Omega_{m0}(1+z)^3f\Omega_{tac}\sqrt{-\omega_{tac}}}{1-\Omega_t},
\end{equation}
where $\rho_{cr,0} = 3M_p^2H_0^2$ is the critical energy density of the universe at the present epoch. In addition, using the relation $\dot{\phi}=-H(1+z)\frac{d\phi}{dz}$, the evolution of the interacting holographic
tachyon scalar field $\phi$  with respect to the redshift z becomes,
\begin{equation}\label{phi}
\frac{1}{H_0^{-1}}\frac{d\phi}{dz}=\pm\frac{\sqrt{1+\omega_{tac}}}{H(1+z)},
\end{equation}
where the sign is arbitrary and can be changed by a redefinition of the field, $\phi \rightarrow -\phi$. Then, by fixing the field amplitude at the present era to be zero, one may obtain the dynamic of the holographic tachyon field. It is difficult to solve equations (\ref{omegaz})-(\ref{phi}) analytically, however, the evolutionary form of the interacting holographic
tachyon field  $\phi$ and $\Omega_{tac}$ can be easily obtained by integrating it numerically
from $z = 0$ to a given value $z$. In addition, from the constructed holographic
tachyon model, the evolution of $V(\phi)$ with respect to $\phi$ can be determined.

In the following numerical computation, we assume that the function $f(\phi)$ behaves exponentially as $f(\phi)=f_0e^{b\phi(z)}$. Since $\dot{f}(\phi)$ is present in the interaction term $Q$, the parameters that directly determine the dynamics of the interaction are $b$ and $f_0$ together with $\rho_m$ and $\gamma$. Note that $f_0=0$ or $b=0$ leads to the absence of the interaction.

 In Fig. 1 and 2, by solving the equations (\ref{omegaz})-(\ref{phi}), the reconstructed scalar field $\phi (z)$ and $V(\phi)$ are shown. The curves are plotted for different values of $b$, $c$, $f_ 0$ and with the fractional matter density selected to be $\Omega_{m,0}=0.27$. From Fig. 2, the reconstructed tachyon potentials are more steep in the early epoch ($z\thicksim 0.5$) and tending to be flat in near future when the scalar field $\phi$ is negative. Therefore, the tachyon scalar field rolls down the potential more slowly with the kinetic energy gradually decreasing in near future where from Fig. 3a the tachyon EoS parameter become tangent to -1 when $\dot{\phi}\rightarrow 0$. This scenario is similar to the holographic quintessence scenario \cite{L}. Note that, from Fig.1, the tachyon scalar field increases with $z$, becomes finite at high redshift and decreases as the universe expands. Note also that, due to smallness of interaction term, the behavior of the holographic scalar field is similar in the presence and absence of the interaction term which is compatible with the observational data \cite{Setare3}.\\

\begin{figure}[t]
\includegraphics[scale=.25]{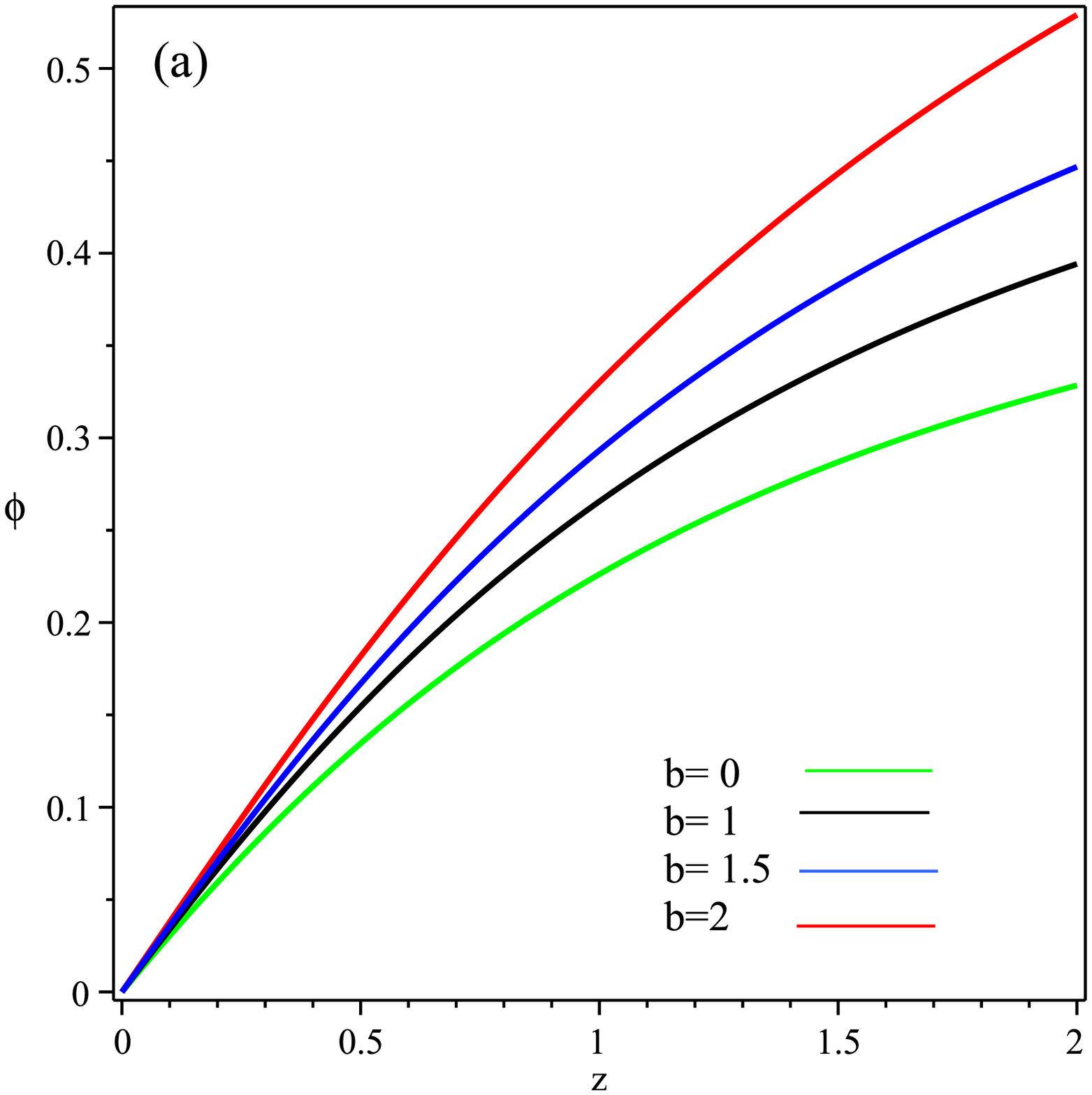}\hspace{0.1 cm}\includegraphics[scale=.25]{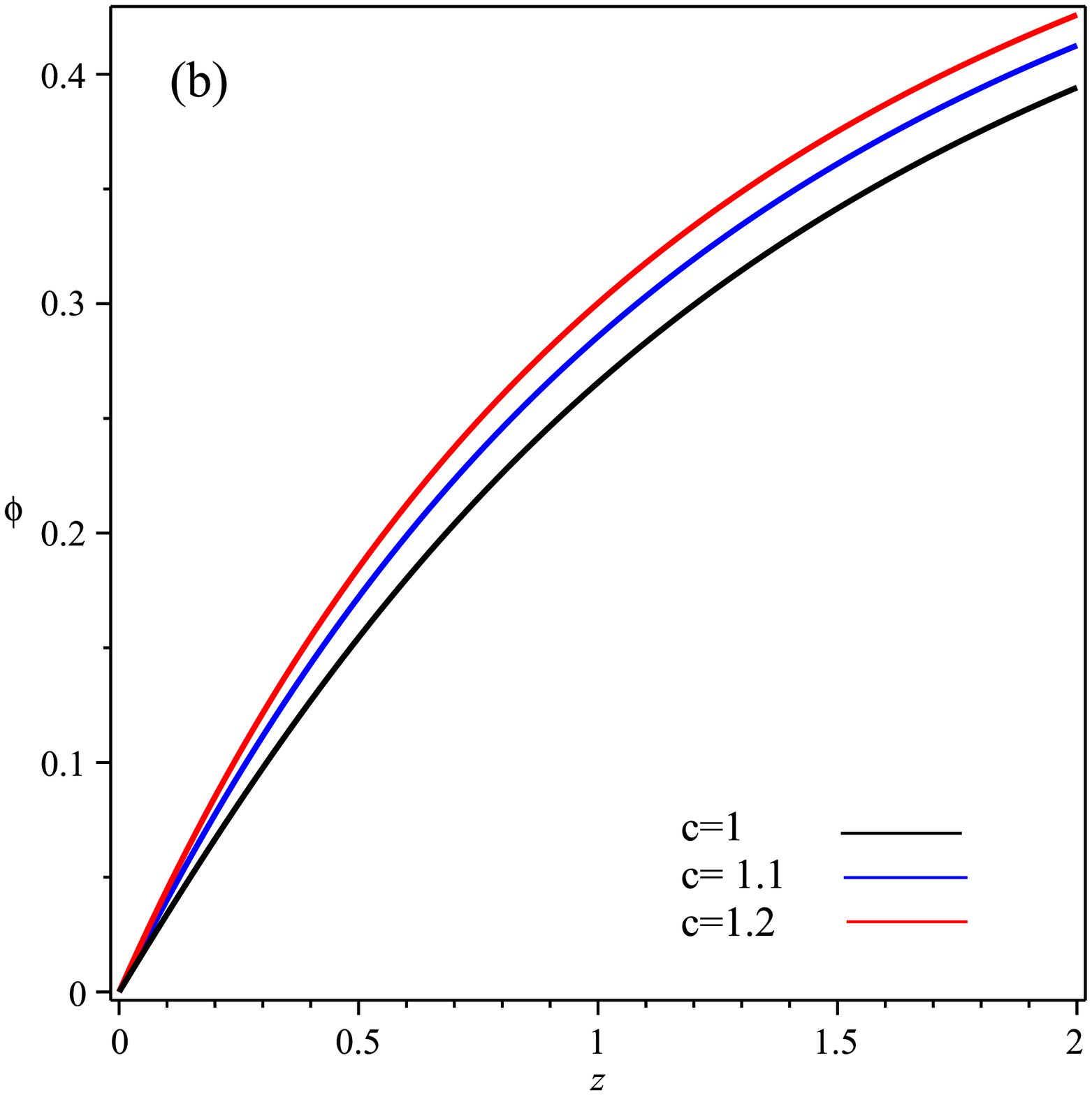}\hspace{0.1 cm}\includegraphics[scale=.25]{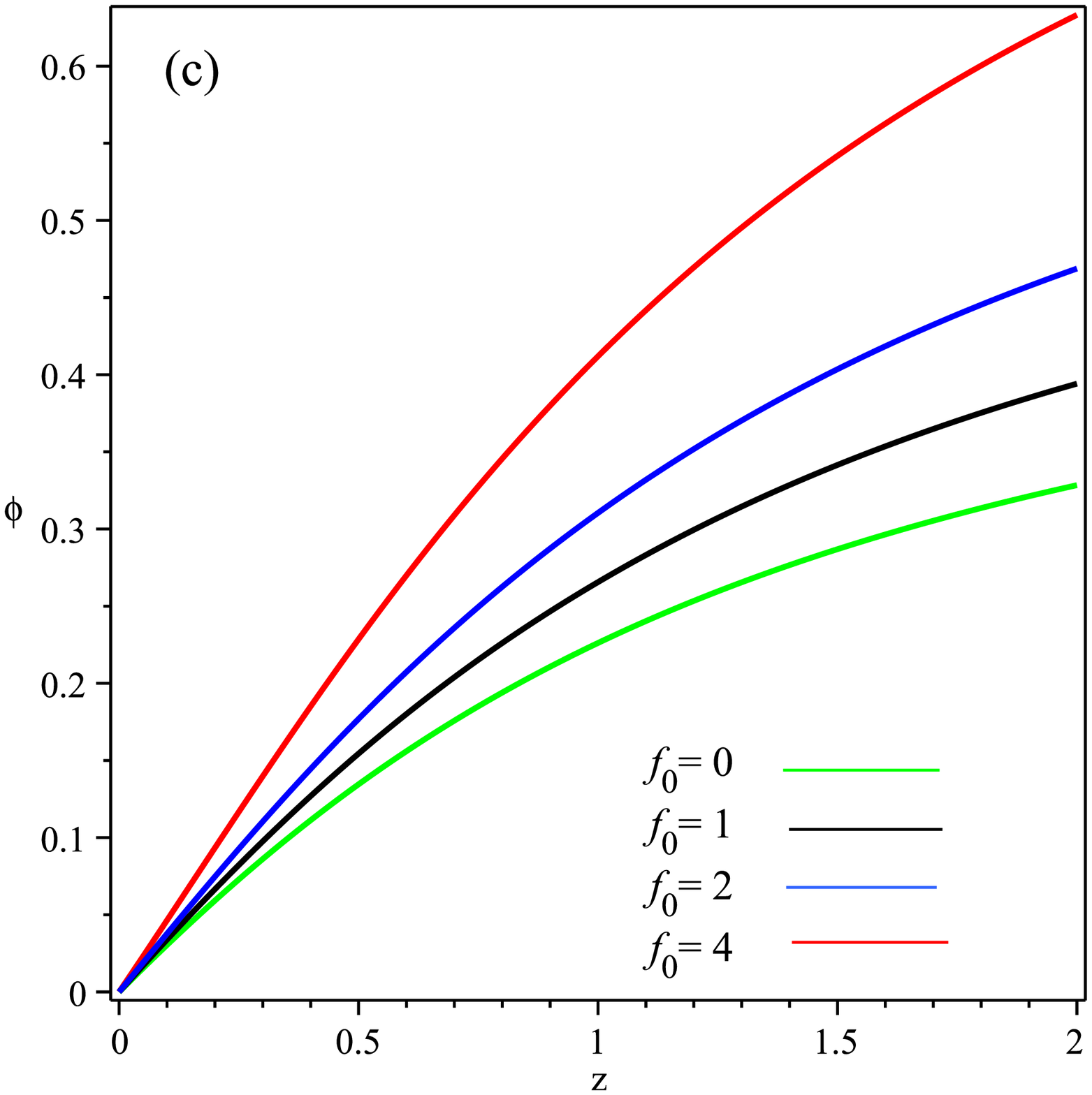}\hspace{0.1 cm}
\caption{ The behavior of the scalar field with respect to $z$,  for different parameters $b$, $c$, $f_ 0$ and $\Omega_{m,0}=0.27$ where $\phi$
is in unit of $H_0^{-1}$.
(a) $c=1, f_0=1$, (b) $b=1, f_0=1$, (c) $c=1, b=1$} 
\end{figure}

\begin{figure}[t]
\includegraphics[scale=.25]{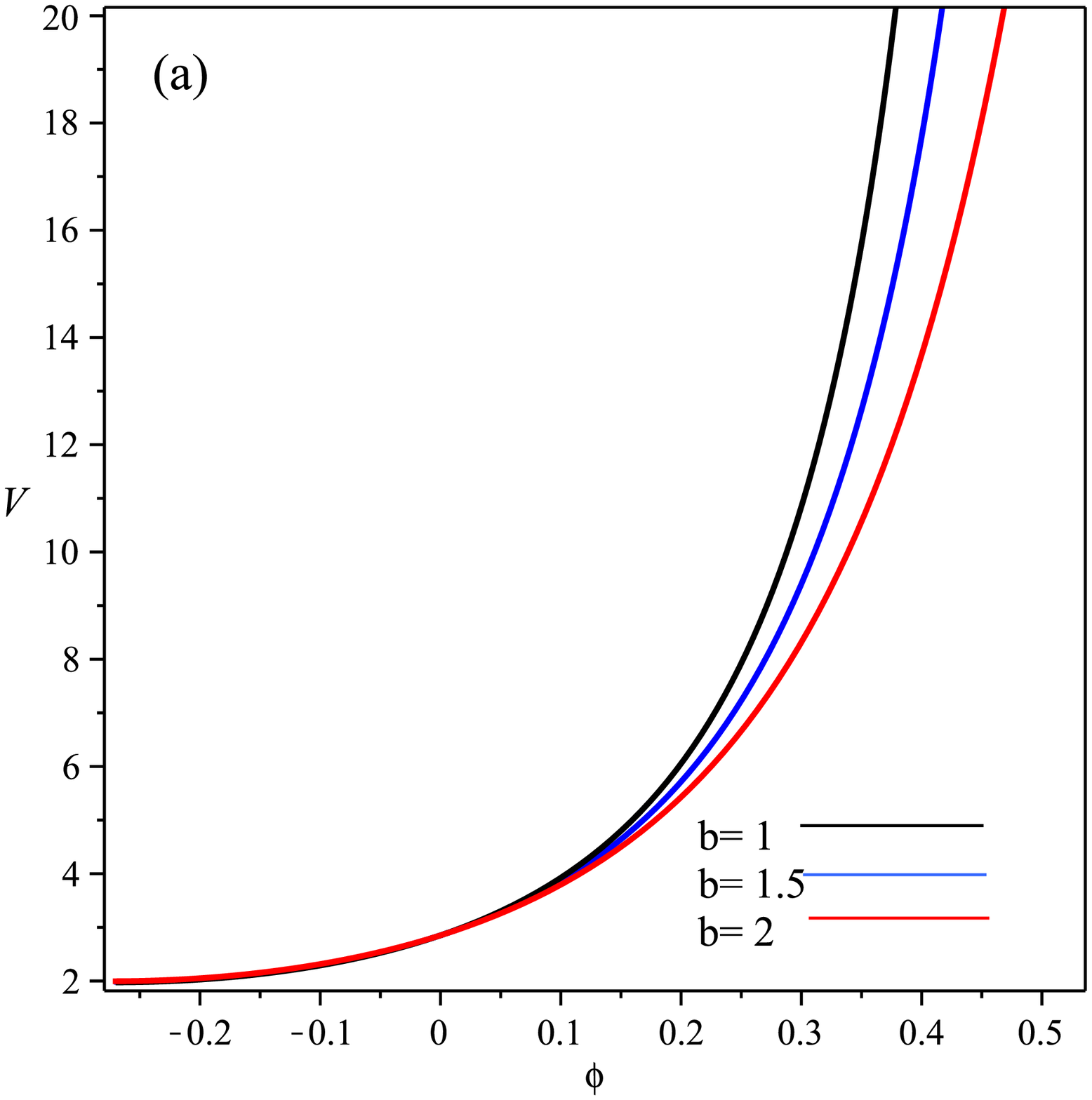}\hspace{0.1 cm}\includegraphics[scale=.25]{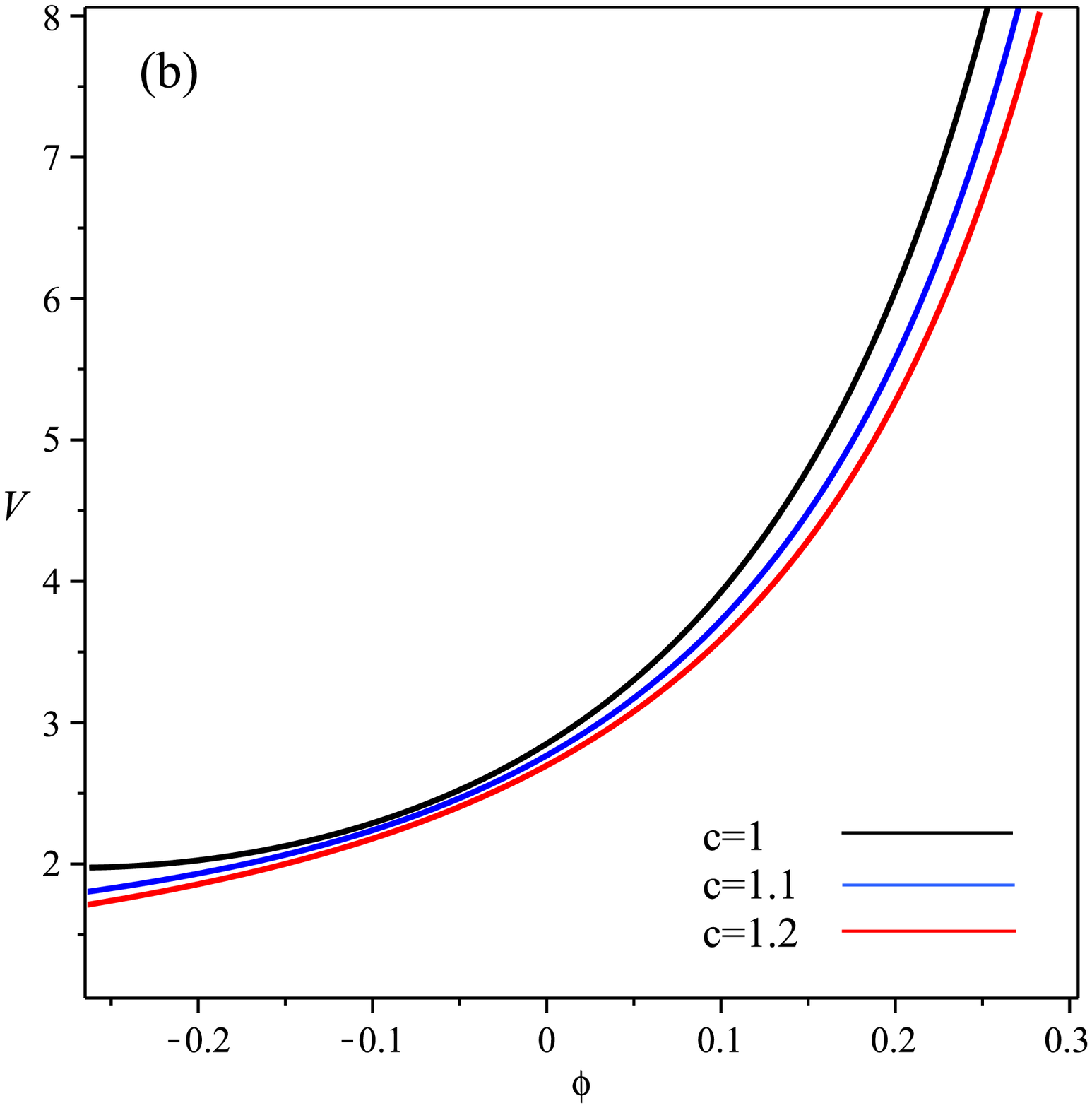}\hspace{0.1 cm}\includegraphics[scale=.25]{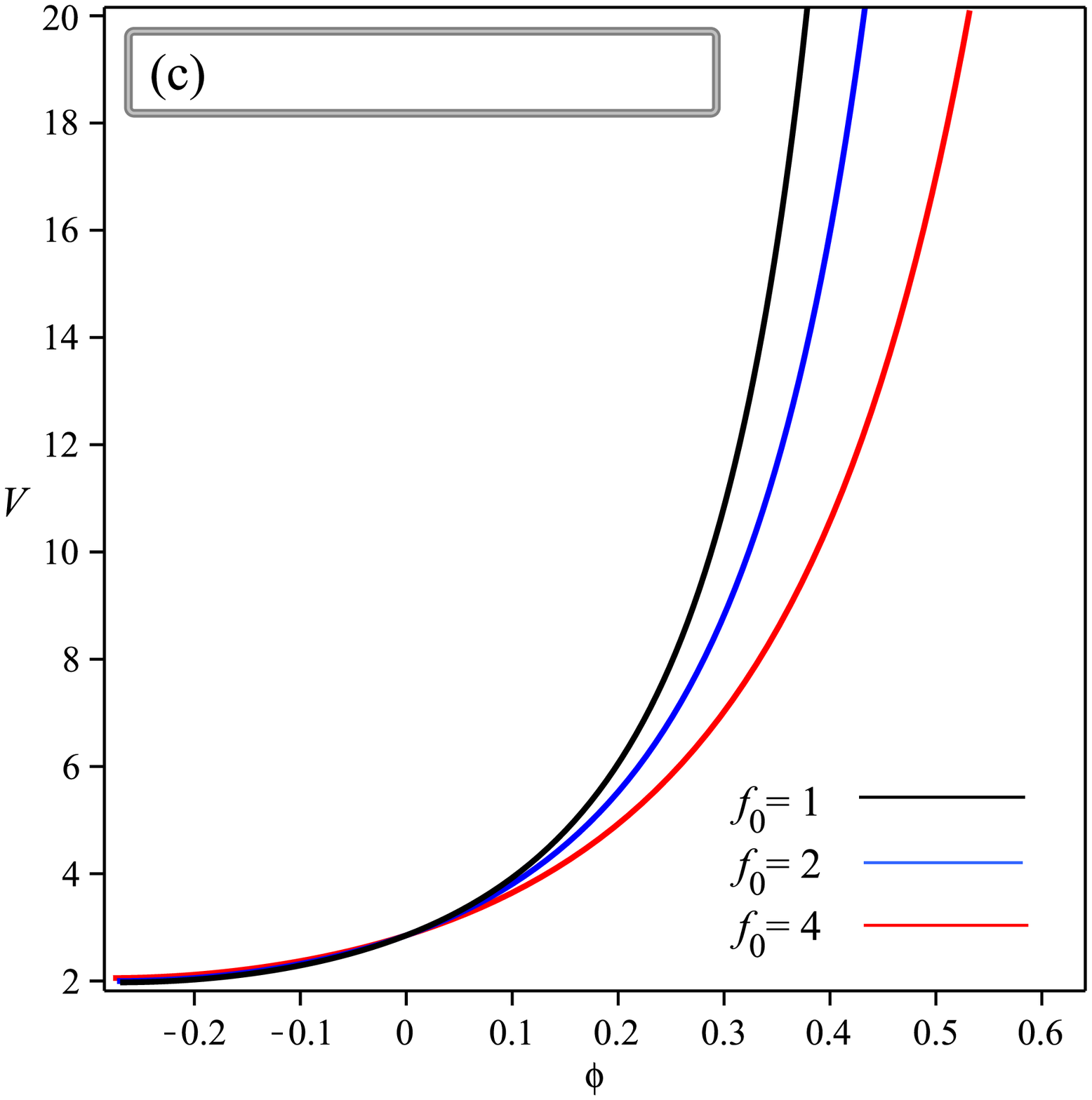}\hspace{0.1 cm}
\caption{ The behavior of the potential for the interacting holographic tachyon model with respect to $\phi$, for different parameters $b$, $c$,
 $f_ 0$ and $\Omega_{m,0}=0.27$ where $\phi$ is in unit of $H_0^{-1}$ and  $V(\phi)$ in $\rho_{cr,0}$.
(a) $c=1, f_0=1$, (b) $b=1, f_0=1$, (c) $c=1, b=1$} 
\end{figure}

In addition, the dynamics of the ratios of the energy densities of the HDE, $\Omega_{tac}$, and DM, $\Omega_{ch}$, to the critical energy density in the universe are shown in Fig.3b. The graph shows that the impact of the interaction between IHDE and DM increases at a faster rate compared to the non-interacting case. It also shows that in the presence of interaction the domination of HDE occurs later in $z$.\\

\begin{figure}[t]
\includegraphics[scale=.35]{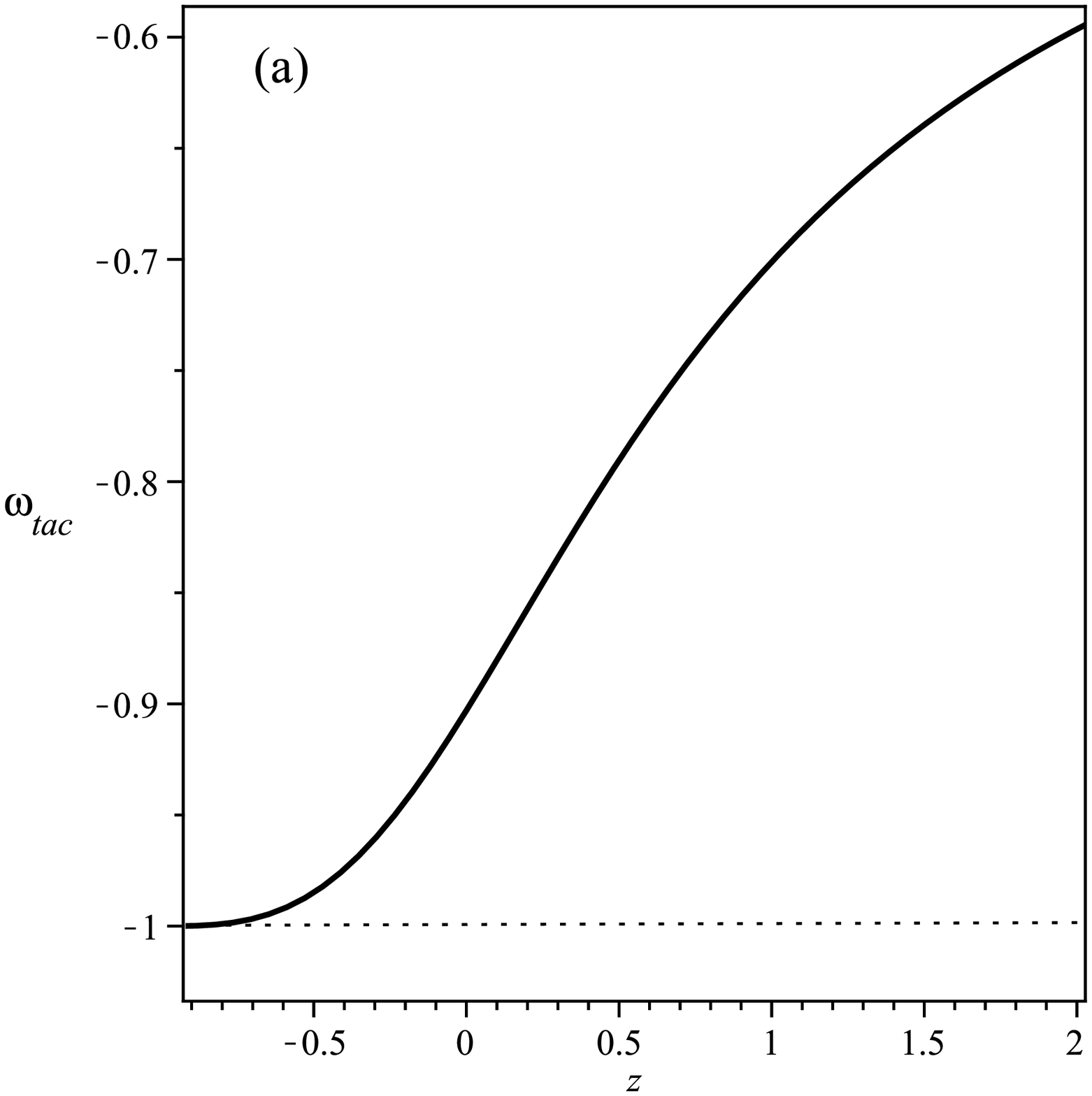}\hspace{0.1 cm}\includegraphics[scale=.35]{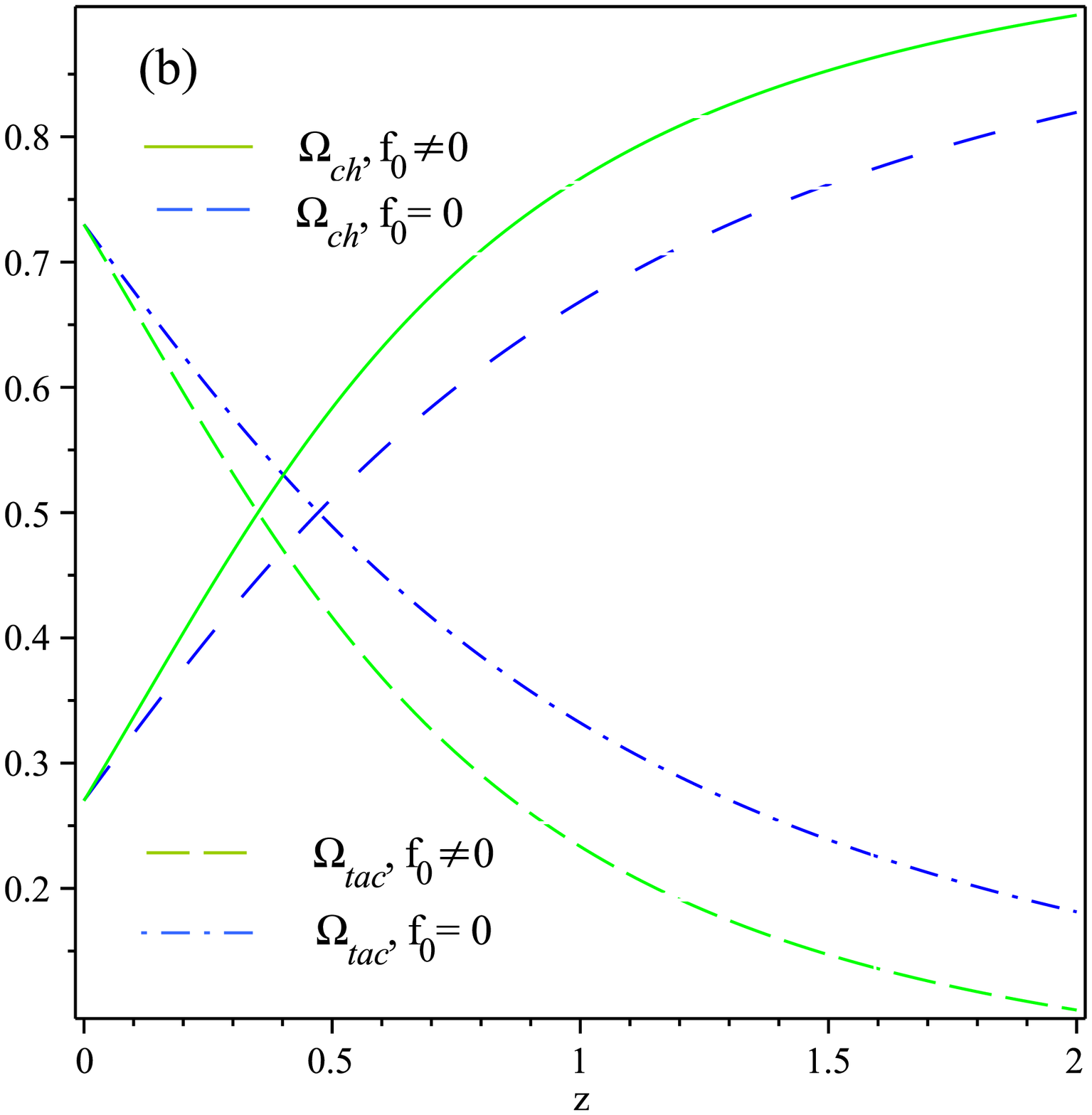}\hspace{0.1 cm}
\caption{Variation of $\Omega_{ch}$  and $\Omega_{tac}$ with respect to redshift $z$ for the holographic tachyon model with and without interaction.
 We put $c=1$ and $\Omega_{tac,0}=0.73$} 
\end{figure}

From the above equations one can easily obtain the total EoS and deceleration parameters $q$ for the model as,
\begin{eqnarray}
\omega_{tot}&=&-\frac{2}{3c}{\Omega_{tac}}^{3/2}-\frac{1}{3}\Omega_{tac}+\frac{Q}{9H^3M_P^2}\label{w}\\
q&=&-\frac{\Omega_{tac}^{3/2}}{c}-\frac{1}{2}\Omega_{tac}+\frac{1}{2}+\frac{Q}{6H^3M_P^2},\label{q}
\end{eqnarray}
In Fig. 4, the deceleration parameter for different parameters $b$, $c$ and $f_ 0$ are shown. Figs. 4a) and 4c), for a fixed $c$, show the effect of interaction term on the acceleration of the universe. From these graphs, since in the interaction the IHDE decays into DM, the cosmic acceleration starts later with interaction compare to the case without interaction since IHDE dominates later.  Also for larger interaction, since more IHDE decays into DM, the acceleration starts later.

\begin{figure}[t]
\includegraphics[scale=.25]{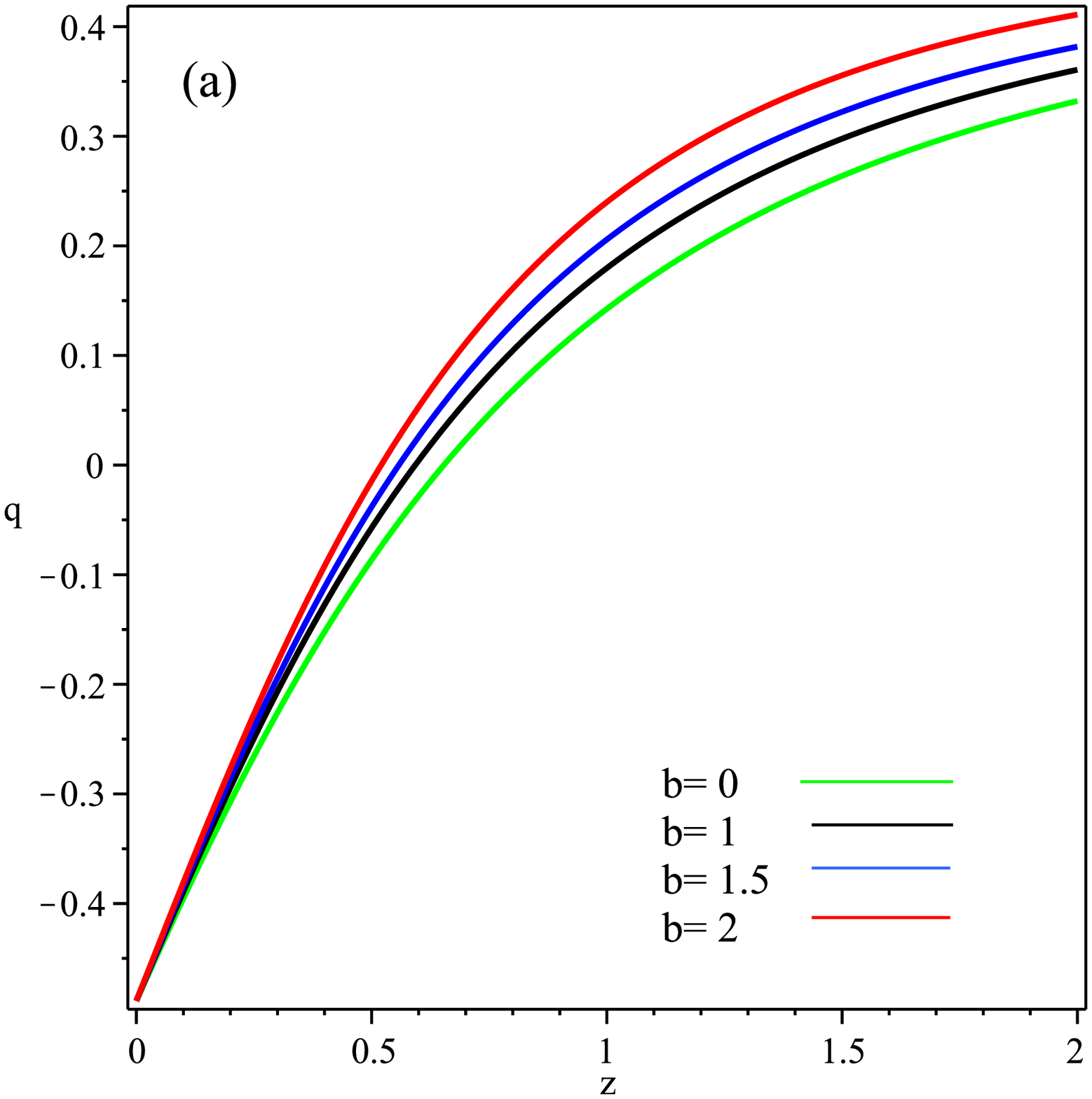} \hspace{0.1 cm}\includegraphics[scale=.25]{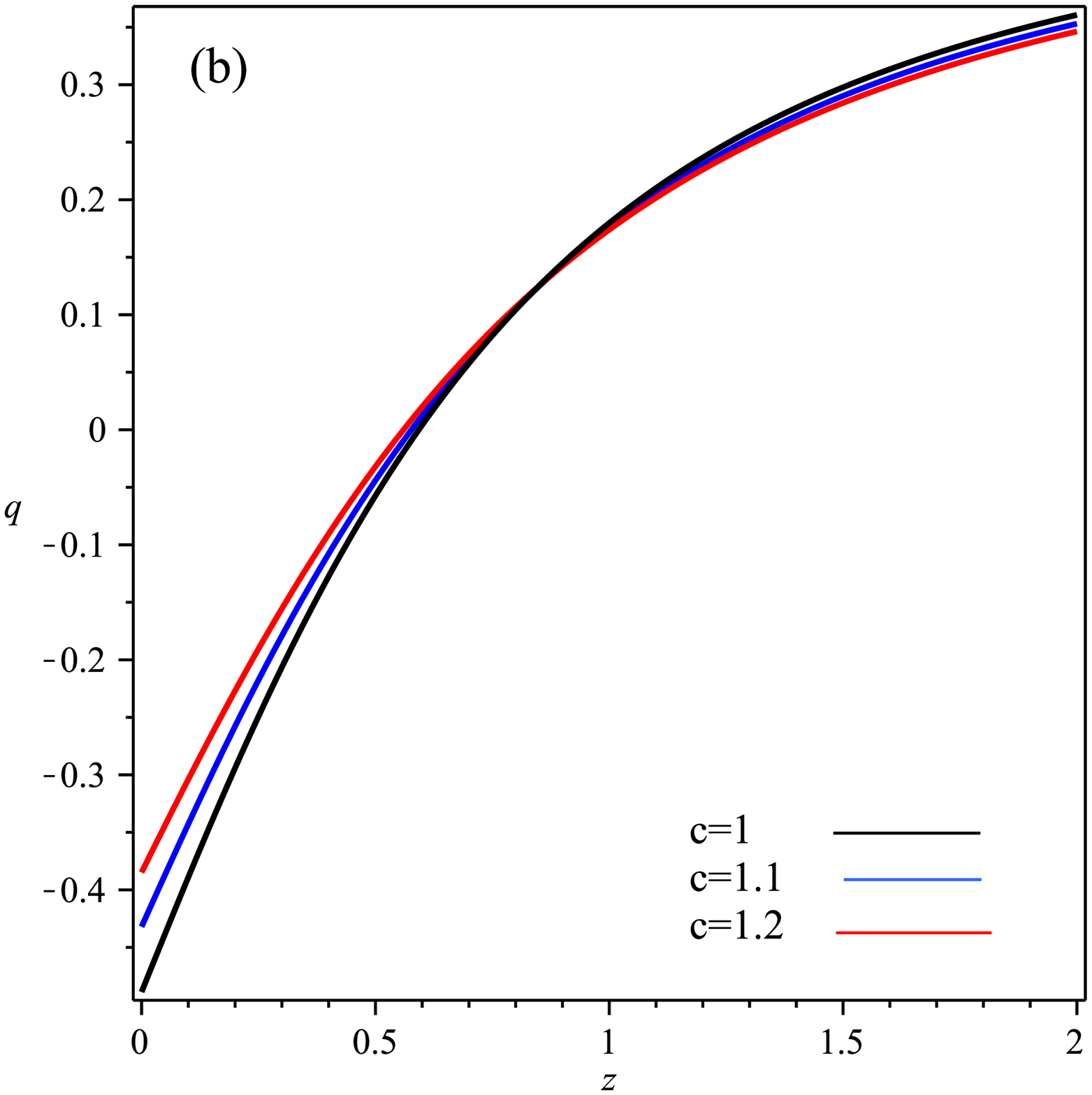} \hspace{0.1 cm}\includegraphics[scale=.25]{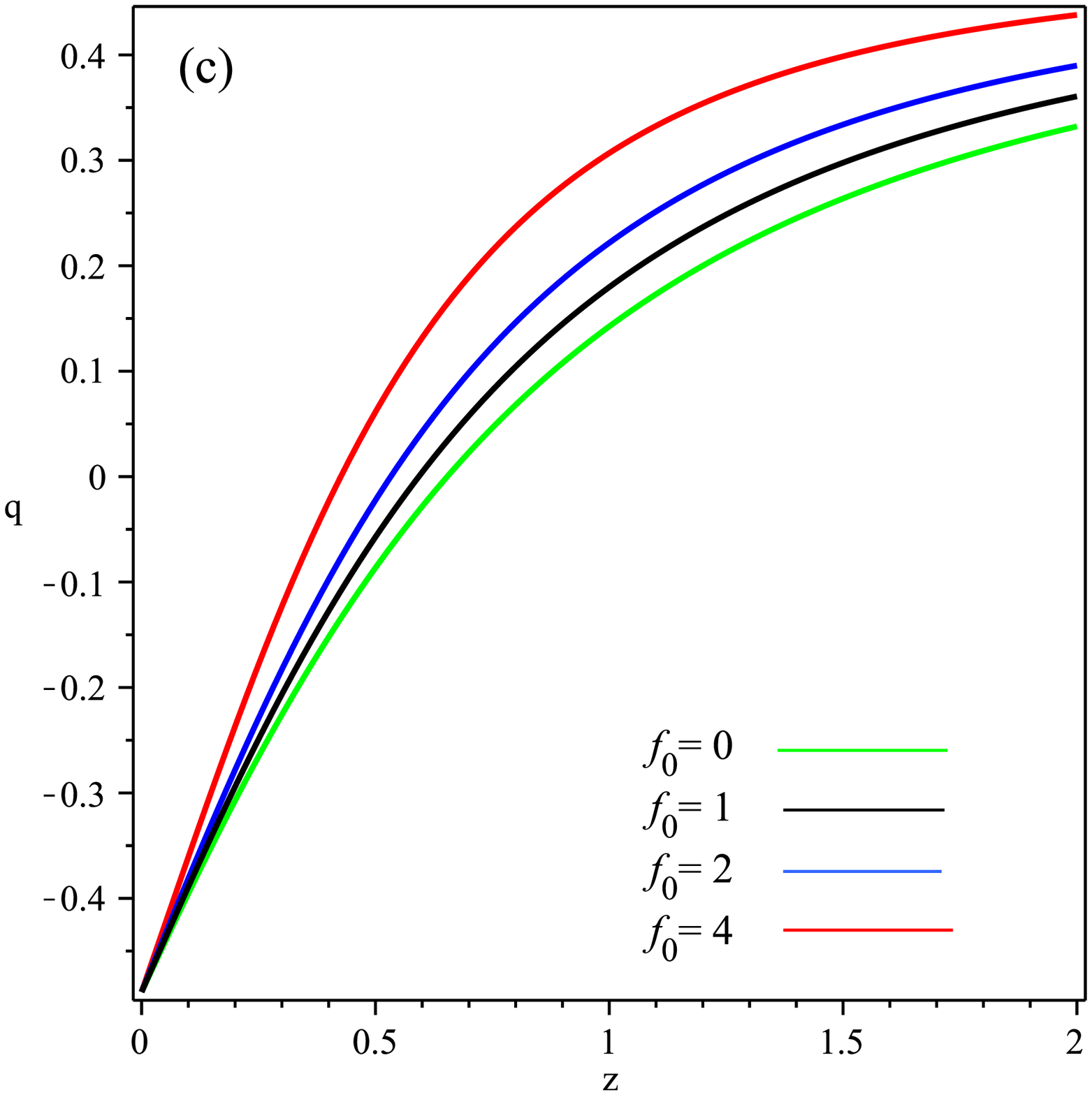} \hspace{0.1 cm}
\caption{The behavior of the deceleration parameter with respect to $z$, for different parameters $b$, $c$, $f_ 0$ and $\Omega_{m,0}=0.27$.
(a) $c=1, f_0=1$, (b) $b=1, f_0=1$, (c) $c=1, b=1$} 
\end{figure}

Noting that in the constructed holographic model, equations (\ref{omegaz})-(\ref{phi}) satisfy the condition  $c\geq 1$. On the other hand, the observational data constrains the $c$ value. However, the latest observational data, including the SNe Ia, the CMB shift parameter given by the WMAP observations, and the BAO measurements from SDSS, show the possibilities of both $c>1$ and $c<1$, and their likelihood are almost equal within $3\sigma$ error range\cite{Setare3}\cite{Liu}.

\section{Cosmological Test}

Luminosity distance quantity, $d_L(z)$, determines DE density from observations.
In our model with the $H(z)$ obtained from numerical calculation, we have
\begin{equation}\label{dl}
d_{L}(z)=(1+z)\int_0^z{\frac{dz'}{H(z')}}=\frac{1+z}{H_0}\int_0^z{\frac{\sqrt{1-\Omega_t}dz'}{\sqrt{\Omega_{m0}(1+z)^3f}}}\cdot
\end{equation}
The difference between the absolute and
apparent luminosity of a distance object called distance modulus, $\mu(z)$, and is given by $\mu(z) = 25 + 5\log_{10}d_L(z)$. In Fig. 5, we
compare our model for the distance modulus with the
observational data. Here, we have used recent observational
data, including Sne Ia which consists of 557 data points belonging to the Union sample\cite{Aman}.\\

\begin{figure}[t]
\includegraphics[scale=0.5]{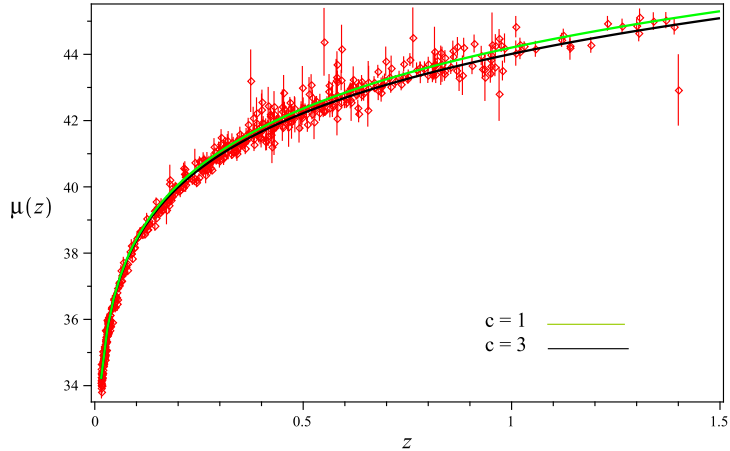}\hspace{0.1 cm}
\caption{Plot of the distance modulus, $\mu(z)$, with respect to the redshift $z$ for two different values of $c=1,3$,
 in comparison with the observational data. We put $f_0=1$ and $b=1$} 
\end{figure}

The graph shows that for $c\geq 1$ the distance modulus obtained from numerical calculation for the model better fits the observational data.

\section{Summary and Remarks}

In this work we have proposed a holographic
model of tachyon-chameleon cosmology by taking into account the holographic principle
of quantum gravity; by implementing a combination of holographic DE model and interacting tachyon chameleon fields. In \cite{Setare3} the holographic dark energy is described by the tachyon field in a certain way. Then a correspondence between the HDE and tachyon model of dark energy has been established in two different cases, flat and non-flat FRW universes. The author has shown that the value of $c$ is a key point to the feature of the HDE and the ultimate fate of the universe as well.

In holographic models, one considers that the scalar field models of DE
are effective theories of an underlying theory of DE. In particular, the tachyon scalar field carries some
holographic feature that with $c \geq 1$ is capable of realizing the holographic
evolution of the universe. In our model, by connection between the tachyon and the holographic
DE, the potential of the holographic tachyon has been constructed. Although the effect of the interaction term is very small,
the possibility of a small, but calculable, interaction
in the dark sector remains open, and future investigations, with more realistic models and
more observational data are necessary to solve this problem \cite{Sandro}. The scalar field in our formalism plays two roles; as a chameleon field interacts with the matter in the universe and as a tachyon field plays the role of DE. In the more familiar models, the interacting term between IHDE and DM is introduced in the formalism based on phenomenological grounds in studying a suitable coupling between quintessence scalar field and pressureless CDM to alleviate the coincidence problem \cite{Wang}. In our work, different from other approaches, the interacting term naturally appears in the model from the interaction between chameleon and matter fields in the universe. The interaction terms is similar to the proposed ones introduced in the IHDE models. A comparison of the model with recent observational data shows that for the parameter $c$ greater than one, the model for distance modulus better fits the observational data.

In \cite{Setare2} and \cite{Setare1} the author considered $L$ measured from the sphere of the horizon, as system's IR cut-off and obtained the equation of state for the interacting holographic energy density in the non-flat universe. Further, in \cite{Setare2} the effects of the interaction between a brane universe and the bulk and in \cite{Setare1}, between holographic energy density and CDM are studied. In our paper, the tachyon field in connection with IHDE shows a quintessence like equation of state parameter in the past that become tangent to the phantom divide line in future. The deceleration parameter in the model shows a transition from deceleration to acceleration at about $z\simeq 0.5$ in the past for different model parameters compatible with recent observational data.

\end{document}